\documentclass[12pt]{article}
\usepackage[margin=0.95 in]{geometry}
\usepackage{amsmath} 
\usepackage{amssymb,amsfonts}
\usepackage[all]{xy}
\usepackage{graphicx}
\usepackage{simplewick}
\usepackage{amsmath}
\usepackage{amssymb}
\usepackage{float}
\usepackage{array}
\usepackage{tikz}
\usepackage{mathtools}
\usepackage{mathrsfs} 
\usepackage[hidelinks]{hyperref}
\usepackage{cite}
\numberwithin{equation}{section}
\usepackage[utf8x]{inputenc}

\newcommand{\be}{\begin{equation}}
\newcommand{\ee}{\end{equation}}
\def\bea{\begin{eqnarray}}
\def\eea{\end{eqnarray}}

\newcommand{\ii}{i}
    \newcommand{\Imag}{\text{Im}}
\newcommand{\pa}{\partial}

\newcommand{\Zint}{Z_{\text{int}}}
\newcommand{\cint}{c_{\text{int}}}

\numberwithin{equation}{section}
\numberwithin{table}{section}

\begin{document}

\hypersetup{pageanchor=false}
\begin{titlepage}
\vbox{
    \halign{#\hfil         \cr
           } 
      }  
\vspace*{15mm}
\begin{center}
{\Large \bf 
Twisted Strings in Three-dimensional Black Holes}

\vspace*{10mm}

{\large Sujay K. Ashok$^a$ and Jan Troost$^b$}
\vspace*{8mm}

$^a$The Institute of Mathematical Sciences, \\
          Homi Bhabha National Institute (HBNI),\\
		 IV Cross Road, C.I.T. Campus, \\
	 Taramani, Chennai, India 600113

\vspace{.8cm}

$^b$ Laboratoire de Physique de l'\'Ecole Normale Sup\'erieure \\ 
 \hskip -.05cm
 CNRS, ENS, Universit\'e PSL,  Sorbonne Universit\'e, \\
 Universit\'e de Paris 
 \hskip -.05cm F-75005 Paris, France	 
\vspace*{0.8cm}
\end{center}

\begin{abstract}
We revisit the classical dynamics of fundamental strings in BTZ black holes with NSNS flux. We  analyze probe strings in the black hole background and determine their classical energy using a Nambu-Goto  action.   Three-dimensional gravity has a non-local effect on the metric surrounding a massive object which can be measured by  winding strings. We also study the behaviour of short and long strings in the Wess-Zumino-Witten orbifold and show that their energy  matches the long string probe analysis. In the process, we fix subtleties in the Lorentzian choice of NSNS flux. In  Euclidean signature, we analyze the one-loop fundamental string partition function on the BTZ black hole and interpret the off-shell degrees of freedom as those of a coset orbifold.  Finally, we speculate on how to identify the quantum Lorentzian string spectrum from the Euclidean orbifold partition function.

\end{abstract}

\end{titlepage}
\hypersetup{pageanchor=true}

\setcounter{tocdepth}{2}
\tableofcontents

\section{Introduction}
Anti-de Sitter space-times play a central role in our study of holography in quantum gravity. Moreover, string theory provides a theory of quantum gravity in which holography is on a firm footing. Thus, understanding string theory in asymptotically anti-de Sitter space-times even better remains a crucial task.
Analyzing the classical dynamics of fundamental strings in three-dimensional anti-de Sitter space-time was instrumental to make headway in the determination of the  spectrum in the quantum theory \cite{Maldacena:2000hw,Maldacena:2000kv}. See e.g. \cite{Petropoulos:1989fc} and references therein for earlier work. This laid the basis for making progress in an example of holography which is controlled even at string scale curvature. It remains one of the best  understood examples of holography in quantum gravity. See e.g. \cite{Eberhardt:2019ywk,Dei:2021yom,Ashok:2021iqn,Martinec:2021vpk,Eberhardt:2021vsx} for further recent progress.

In this paper, we wish to make a contribution to our understanding of string theory in other asymptotically $AdS_3$ space-times. Prominent examples of such backgrounds are provided by black hole solutions in three-dimensional anti-de Sitter space-time, the BTZ black holes \cite{Banados:1992wn,Banados:1992gq}. While many results on the classical solutions of fundamental strings moving in these black holes have been obtained, as well as some insight in the quantum spectrum \cite{Natsuume:1996ij,Hemming:2001we,Troost:2002wk,Hemming:2002kd,Rangamani:2007fz,Berkooz:2007fe,Mertens:2014nca,Mertens:2015ola}, we believe they are usefully extended by our results. Indeed, any further step towards understanding a quantum theory of gravity on a black hole background is worthwhile.

In section \ref{Definition}, we review the Lorentzian BTZ black hole background, its topology as well as its relation to the $SL(2,\mathbb{R})$ group manifold. We will recognize that string world sheet embeddings in the BTZ space-time give rise to interesting questions as to the definition of the string world sheet two-form potential term. In section  \ref{Classical} we review classical string solutions to the extent that they were known, based on particle geodesics with added winding.
We embed the solutions in the Wess-Zumino-Witten orbifold representing the BTZ black hole and analyze their energy in more detail. In section \ref{LongStringProbe}, we perform a Nambu-Goto analysis of a long string winding the black hole.
We relate the analysis to that of a long string in a generic asymptotically $AdS_3$ space-time with curved boundary. Comparing the analysis to the Wess-Zumino-Witten results allows us to identify a preferred definition of the NSNS potential term in the world sheet model. Section \ref{PartitionFunction} is dedicated to a Euclidean counterpart to the preceding sections. We analyze how to wind Euclidean solutions such that one obtains classical solutions in the twisted sector of the black hole orbifold. The path integral in the Euclidean orbifold coset conformal field theory is computed, and we identify expected contributions to the partition sum. Section \ref{Conclusions} contains conclusions while  Appendix \ref{Geometries} briefly reviews relevant geometries. Appendix \ref{generalprobe} ties up a loose end.

\section{String Theory on  Black Hole Backgrounds}
\label{STLBH}
\label{Definition}
\label{Model}
In this section, we review the topology of the BTZ black hole and the relation of the black hole to the $SL(2,\mathbb{R})$ group manifold \cite{Banados:1992wn,Banados:1992gq}. We stress that the topology of the black hole leads to subtleties in the definition of the perturbative fundamental string world sheet model. 
\subsection{The BTZ Black Hole}
We study fundamental string theory on the BTZ black hole background \cite{Banados:1992wn,Banados:1992gq} with metric
\be
ds^2 = \frac{l^2 r^2}{(r^2-r_+^2)(r^2-r_-^2)}dr^2 - \frac{(r^2-r_+^2)(r^2-r_-^2)}{l^2 r^2}dt^2+r^2(d\phi - \frac{r_+r_-}{l r^2} dt)^2~ \, ,
\label{BTZMetric}
\ee
supplemented with an NSNS flux $H_{(3)}$ proportional to the volume of space-time.  We imagine a number of extra dimensions necessary to form a consistent bosonic or super string background \cite{Polchinski:1998rq}. The inner horizon at radius $r_-$ and the outer horizon at radius $r_+$ are set by the black hole mass $M$ and angular momentum $J$ through:\footnote{Our convention is that the radius $r_-$ is a positive real number. 
The mass $M$ is dimensionless while the angular momentum $J$ has dimension of length. The Ricci scalar equals $R=-6/l^2$.}${}^{,}$\footnote{ Note that if we make the coordinates $r,t$ dimensionless through the replacement $(r,t) \rightarrow (lr,lt)$, then we find a factor of $l^2$ up front in the metric. We will set this length scale equal to $l=\sqrt{k \alpha'}$ later on. }
\begin{equation}
M = \frac{r_+^2+r_-^2}{l^2}~, \qquad J  = \pm \frac{2 r_+ r_-}{l}  \, .
\end{equation}
We will often set the cosmological constant length scale $l$ to be equal to one by a choice of units. The coordinate system (\ref{BTZMetric}) in which the BTZ black hole is  represented makes for an analogy to common  coordinate choices for four-dimensional black holes. However, the geometry of BTZ black holes is quite different.

 \subsection{The Black Hole and the Group}
 As discussed in great detail by BHTZ  \cite{Banados:1992gq}, the BTZ black hole arises from the $SL(2,\mathbb{R})$ group manifold as follows. The $SL(2,\mathbb{R})$ group manifold with left-right invariant metric is the 
three-dimensional hyperboloid
\begin{equation}
x_{-1}^2 + x_{0}^2 -x_1^2 -x_2^2 = 1 \, ,
\end{equation}
with the metric induced from the flat embedding space.
Topologically, there is a circle in the $(x_{-1},x_0)$ plane that can never shrink to zero size. Indeed,  the first homotopy group of the group manifold is non-zero: $\Pi_1(SL(2,\mathbb{R})=\mathbb{Z}$.
The universal covering group {\small $ \widetilde{SL(2,\mathbb{R})}$} covers the group manifold and has a trivial fundamental group. It  has the  topology of $\mathbb{R}^3$. 
The BTZ black hole is obtained by dividing the universal covering group by a $\mathbb{Z}$ orbifold action. A generator of the group acts on a $2 \times 2$ standard matrix representation $g$ of the group elements of $SL(2,\mathbb{R})$ as:
\begin{equation}
g \rightarrow  e^{\pi (r_+-r_-) \sigma_3} \, g \, e^{\pi (r_++r_-) \sigma_3} \, ,
\end{equation}
where $\sigma_3$ is the diagonal Pauli matrix with entries $\pm 1$.
Crucially, when both the horizon radii $r_+$ and $r_-$ are non-zero, the orbifold action has no fixed points. This is equally true on the covering group. Therefore, the orbifolded manifold is regular and has the topology of $\mathbb{R}^2 \times S^1$. The circle generated by the orbifold action is the one parameterized by the coordinate $\phi$ in the patch where the BTZ coordinates (\ref{BTZMetric}) are valid. 

The BHTZ analysis carries further  and involves  the metric inherited from the group manifold and the causal structure it induces \cite{Banados:1992gq}. They exclude from the BTZ geometry the region of the manifold which contains circles that are null or time-like on the physical ground that these lead to paradoxes, and that the coupling to matter would create a singularity at the boundary surface $r=0$. Wherever the topologically non-trivial circle becomes null, the BHTZ proposal is to introduce a two-dimensional boundary to the BTZ space-time \cite{Banados:1992gq}.

\subsection{An Ambiguity}
We have signaled two aspects of the Lorentzian BTZ black hole backgrounds that make it more subtle to define a unique world sheet theory. The first is the non-trivial topology of the orbifolded universal covering group which is $\mathbb{R}^2 \times S^1$. The second is the introduction of boundaries in space-time. We wish to address subtleties associated to the first aspect and will hardly remark on the second. 

The world sheet theory depends on a NSNS two-form flux potential $B_{(2)}$. One way to define the two-form flux term is as the integral of the field strength $H_{(3)}=dB_{(2)}$ over a three-cycle $\Sigma_3$ with a boundary that coincides with the fundamental string world sheet $\Sigma=\partial \Sigma_3$. If the world sheet $\Sigma$ is topologically trivial (and we have a quantized integral over three-cycles in the target space) then that definition of the flux term in the world sheet action is unique. 
However, we will  study winding strings on BTZ black hole backgrounds. These have world sheets that are rather of the form $\Sigma = \mathbb{R} \times S^1$ where the $S^1$ winds the topologically non-trivial circle in the background $\mathbb{R}^2 \times S^1$ space-time. If the flux term were to fall off at infinity, we could still define the world sheet flux term in terms of the integral of the flux $H$ over a semi-infinite cylinder. However, this fall-off condition is not satisfied in the background under consideration -- it is easy to see that the $H_{(3)}$ flux grows with radius since it is proportional to $\sqrt{-G}$ where $G$ is the determinant of the BTZ metric (\ref{BTZMetric}).
Another attempt to define the theory could go through the compactification of a cylindrical world sheet to a sphere, but this would lead to a spherical world sheet in a space-time of the form $\mathbb{R} \times S^2$ and would therefore give rise to the necessity to choose a gauge invariant flux integral $\int_{S^2} B$ in order to define the model. Thus, we are stuck with string world sheets that do not arise as boundaries of compact three-cycles and thus the action is hard to define in terms of the gauge invariant three-form field strength $H_{(3)}$ only.

We can take the point of view that  world sheets that  have no boundaries have an action equal to the $B_{(2)}$ field integrated over the world sheet $\Sigma$. Because the world sheet has no boundary, the action is  gauge invariant. We argue that this point of view gives rise to a related ambiguity.
To make this discussion concrete, let us focus on the BTZ black hole background in the coordinates (\ref{BTZMetric}). The $H_{(3)}$ flux is proportional to 
\begin{equation}
H_{(3)} = r dr \wedge dt \wedge d \phi \, .
\end{equation}
We can integrate the $H_{(3)}$ flux  up to a $B_{(2)}$ potential:
\begin{equation}
B_{(2)}  = \frac{1}{2} (r^2 -c) dt \wedge d \phi \, ,
\end{equation}
where we introduced a radial integration constant $c$.
Suppose we have a world sheet of topology  $\mathbb{R} \times S^1$ stretching along the directions $(t,\phi)$. Imagine that we wish to define the world sheet action using the integral over a three-cycle which is a finite cylinder inside the space-time topology $\mathbb{R}^2 \times S^1$. The space-time cylinder has one boundary equal to the world sheet that we study, and a second boundary which we will call the cut-off surface $\Sigma_{co}$.  When we integrate the $H_{(3)}$ flux over the finite cylinder, we schematically find:
\begin{align}
\int_{\text{cylinder}} H_{(3)} &= \int_{\Sigma} B_{(2)}- \int_{\Sigma_{co}} B_{(2)}
\nonumber \\
& =  \frac{1}{2} \int_{\Sigma} (r^2-c) dt d \phi
-  \frac{1}{2}\int_{\Sigma_{co}} (r_{co}^2-c) dt d \phi
\nonumber \\
& =   \frac{1}{2} \int (r^2 - r_{co}^2) d t d \phi \, .
\end{align}
What this calculation illustrates is that the ambiguity in the choice of cut-off surface $r_{co}$ is directly related to the ambiguity in our choice of integration constant $c$ in the isolated world sheet $B_{(2)}$-field term.
Therefore, either we need a cut-off surface when we work in terms of the $H_{(3)}$-flux (and the world sheet action is then independent of the constant term in the $B_{(2)}$-field), or we define the world sheet action in terms of the $B_{(2)}$-field in which case it explicitly depends on the constant $c$ (namely the $B_{(2)}$-field flux along the non-trivial two-cycle). 
Thus, the definition of the model requires a choice. We will encounter this subtlety in various guises in what follows  -- here we have concentrated on the bare-bones origin. For related but different discussions of the subtlety, see
e.g. section 5 of \cite{Hemming:2002kd} and section 3 of \cite{Rangamani:2007fz}.

Finally, the potential introduction  of space-time boundaries (at $r=0$) \cite{Banados:1992gq} is a daunting prospect from the view point of the  world sheet theory \cite{Troost:2002wk}. We rather imagine that a  microscopic definition of fundamental string perturbation theory in F1-NS5 backgrounds, with the addition of appropriate string theory matter and sources will take care of the protection of chronology near the BHTZ $r=0$ surface. This deserves to be the object of further study.

\section{Classical Strings in the Black Hole Orbifold}
\label{Classical}
After these preliminary observations, we recall the shape of space-like and time-like geodesics in BTZ black hole backgrounds
\cite{Cruz:1994ir,Troost:2002wk}. We can dress these geodesic solutions  with winding to find winding classical strings in black hole backgrounds \cite{Natsuume:1996ij,Hemming:2001we,Troost:2002wk}. We calculate the on-shell energy generically, and at zero mass in particular. We compare our results to the analysis of classical strings in three-dimensional anti-de Sitter space \cite{Maldacena:2000hw}. These results will provide useful points of reference both to analyze the Nambu-Goto action of a long string probe in section \ref{NambuGoto} and for the analysis of the conformal field theory partition function on the black hole background in section \ref{PartitionFunction}.

\subsection{The Geodesics}
\label{geodesicsubsection}
\label{Geodesics}
To study the motion of classical fundamental strings in the  black hole background, it is useful to understand the motion of particles first \cite{Cruz:1994ir,Troost:2002wk}. 
A relativistic particle with mass $m$ follows geodesics on the black hole. We parameterize the geodesics using world line proper time $\tau$ such that it satisfies the equation:
\begin{equation}
G_{\mu \nu} \frac{d x^\mu}{d \tau} \frac{d x^\nu}{d \tau} 
= -  l^2 m^2   \, .
\end{equation}
The geodesics have been computed in the literature, and we briefly recall the results \cite{Cruz:1994ir,Troost:2002wk}. The energy $E$ and angular momentum $L$ of the particle are two constants of motion associated to the Killing vectors that translate time $t$ and the angular coordinate $\phi$ in the black hole background (\ref{BTZMetric}). The Killing vectors give rise to the conserved quantities:
\begin{align}
E &= (−M +r^2)\dot{t}+\frac{J}{2} \dot{\phi}~,
\nonumber \\
L &= r^2 \dot{\phi}− \frac{J}{2} \dot{t}
\label{AngularMomentumParticle}
\, ,
\end{align}
where we denoted by a dot the derivative with respect to world line time $\tau$. 
There is one remaining differential equation in the radial variable $y=r^2$. In terms of the parameters $(\alpha,\beta)$
\begin{align}
\alpha &= E^2-L^2 +M m^2~, \nonumber \\
\beta &=L^2 M -\frac{m^2J^2}{4}-JEL \, ,
\label{alphabetadefn}
\end{align}
the differential equation reads
\begin{align}
\label{rdotequation}
\dot{y}^2 &= -4 m^2 y^2+4 \alpha y +4 \beta \, .
\end{align}
Time-like, light-like and space-like geodesics can  be integrated from this equation in terms of their energy $E$ and angular momentum $L$. We concentrate on two classes of geodesics that are of particular interest to us. These two classes are  time-like geodesics and  space-like geodesics with large enough energy such that $\alpha^2 + 4 m^2 \beta >0$.
Time-like geodesics are the starting point for finding short winding strings \cite{Maldacena:2000hw}. Time-like geodesics have $m^2>0$ and the solution to the radial equation reads:
\begin{align}
\label{TLgeo}
r^2 &= \frac{\alpha}{2m^2} + \frac{\sqrt{\alpha^2+4 m^2 \beta}}{2m^2} \sin 2m \tau \, .
\end{align} The geodesic stretches over an interval in the radial direction. The minimal and maximal radius depend on the parameters $(M,J;E,L)$. An example geodesic in a $J=0$ zero angular momentum black hole at zero particle angular momentum $L=0$ will have a radial evolution between $r=0$ and $r_{\text {max}}^2=E^2/m^2+r_+^2$, namely from the center to outside the outer horizon.

Space-like geodesics with large enough energy  such that $\alpha^2 + 4 m^2 \beta >0$ are useful to find long winding strings. They have the radial trajectory:
\begin{align}
r^2 &= \frac{\alpha}{2m^2} + \sqrt{\frac{\alpha^2+4 m^2 \beta}{4m^4}} \cosh \sqrt{-4m^2} \tau \, .
\label{SpaceLikeGeodesic}
\end{align}
At zero angular momenta $J=0=L$ and large energy, the radial geodesic for a tachyon stretches from zero radius all the way to infinity. 
For a more complete discussion of classes of geodesic solutions, we refer to \cite{Cruz:1994ir,Troost:2002wk}.

In string theory (or e.g. a Kaluza-Klein theory), the particle on the black hole background has a mass squared determined by the on-shell condition which schematically reads:
\begin{equation}
-  \frac{k  m^2 }{4} +  h = 0 \, ,
\end{equation}
where $h$ is a quantity that measures a property of the internal degree of freedom\footnote{For example, the quantity $h$ could be set by an internal conformal dimension in a compact conformal field theory defining a compact factor of a string background or a mass due to a Kaluza-Klein momentum.} and we have restored the cosmological length scale $l=\sqrt{k \alpha'}$, related to the level $k$ of the Wess-Zumino-Witten model to be discussed in subsection \ref{WZW}.

\subsection{The Winding Strings}
\label{windstringssec}
It is  well-known how to add winding to the string consistently with the string equations of motion in conformal gauge \cite{Maldacena:2000hw, Natsuume:1996ij, Hemming:2002kd,Troost:2002wk}. The winding   solutions read:
\begin{align}
t(\tau,\sigma) & = t_{\text{geodesic}}(\tau) + w \tau 
\nonumber \\
\phi(\tau,\sigma) &= \phi_{\text{geodesic}}(\tau) + w \sigma
\nonumber \\
r(\tau) &= r_{\text{geodesic}} (\tau) \, ,
\label{ClassicalSolutionWithWinding}
\end{align}
where the lower index  indicates any solution for the time evolution of a geodesic as discussed in subsection \ref{Geodesics}. The winding solution is a puffed up geodesic, which winds the angular direction $\phi$ and is appropriately stretched in time. 

We presented the classical solutions in conformal gauge.  To  find the physical spectrum of winding strings, we still need to solve the on-shell conditions. In  the next subsection, we will solve these conditions,  carefully embed the solutions into an orbifold Wess-Zumino-Witten model and analyze their physical properties  in more depth.

\subsection{The Black Hole  Orbifold}

\label{WZW}
In the following subsections, we discuss properties of the  orbifold  Wess-Zumino-Witten model that represents the conformal field theory with the BTZ black hole target. We refer to \cite{Natsuume:1996ij,Hemming:2001we,Troost:2002wk,Hemming:2002kd,Rangamani:2007fz,Berkooz:2007fe} for related analyses. Compared to these useful resources, in this section we present in more detail properties of the classical solutions and solve more explicitly for the on-shell energy.

In our study we concentrate on the region outside the outer horizon. That region  can be covered in terms of a group element parameterization:\footnote{Implicitly, we always take the universal cover $\widetilde{SL(2,\mathbb{R})}$. 
}
\begin{align}
g &= 
e^{\frac{r_+-r_-}{2} (t+\phi) \sigma_3} \, e^{\rho \sigma_1} \, e^{-\frac{r_++r_-}{2} (t-\phi) \sigma_3}~,
\end{align}
which consists of three factors equal to exponentials of hyperbolic generators of $sl(2,\mathbb{R})$.
The radial coordinate $\rho$ is related to the BTZ radial coordinate $r$ through
\begin{align}
\cosh^2 \rho &= \frac{r^2-r_-^2}{r_+^2-r_-^2}
\, .
\end{align}
We list the (most relevant) current components $J,\bar{J}$ in the $\sigma_3$ direction as well as the energy-momentum tensor in black hole coordinates -- with $x^{\pm} = \tau \pm \sigma$  --:
\begin{align}
 J &= -\frac{k}{r_+-r_-} \left( (r^2-r_+ r_-) \partial_+ \phi +(-r^2+r_+^2+r_-^2-r_+ r_-) \partial_+ t \right) \nonumber 
 \\
 \bar{J}  &= \frac{k}{r_++r_-} \left( (r^2+r_+ r_-) \partial_- \phi +(r^2-r_+^2-r_-^2-r_+ r_-) \partial_- t \right) 
\label{CurrentComponents} \\
T 
&= k \Big( r^2\, (\pa_+\phi)^2+ \, (r_+^2+r_-^2 - r^2)\, (\pa_+ t)^2 - 2\, \, r_+\, r_-\, \pa_+ t\, \pa_+\phi \cr
& \hspace{5.7cm} + \, \frac{ r^2}{(r^2-r_+^2)(r^2-r_-^2)}\, (\pa_+ r)^2 \Big) ~.
\end{align}
A similar expression holds for the left moving stress tensor component with $\pa_+$ replaced by $\pa_-$. As discussed in section \ref{Definition}, 
the black hole orbifold corresponds to a $\mathbb{Z}$ action on the universal cover $\widetilde{G}$ of the group $G=SL(2,\mathbb{R})$. The generator of the group is embedded in the left-right isometry group $\widetilde{G}_L \times \widetilde{G}_R$ that acts on the universal cover. The group action is generated by the zero modes of the left and right currents. We recall that one chooses an embedding of the group $\mathbb{Z}$ such that the action of a generator of the group  is \cite{Banados:1992gq}:
\begin{equation}
g \longrightarrow  e^{\pi (r_+-r_-) \sigma_3} \, g \, e^{\pi (r_++r_-) \sigma_3} \, ,
\end{equation}
which boils down to the identification $\phi \equiv \phi + 2 \pi$. 
Thus, the action of the group $\widetilde{G}_L \times \widetilde{G}_R$ that commutes with the orbifold action consists of the (left and right) one parameter hyperbolic group actions which are obtained by exponentiating the generator $\sigma_3$. These unbroken symmetries give rise to the energy and angular momentum generators. 

We can relate these formulas to those obtained in subsection \ref{geodesicsubsection}. For instance, we can  evaluate the currents and stress tensor on the geodesic solutions. If we declare that all space-time coordinates are $\sigma$ independent, we can  substitute the expressions for the time derivatives $\dot{t}$ and $\dot\phi$ (see equation  (\ref{AngularMomentumParticle})), and $\dot r$ (see equation \eqref{rdotequation}), which are all given in terms of the conserved particle energy $E$ and angular momentum $L$ in the formulas (\ref{CurrentComponents}) for the currents and stress tensor.  We find the relations valid for particle geodesics: 
\begin{align}
\label{JTparticle}
{J}_{\text{particle}}
&= \frac{k}{r_+-r_-} \frac{E-L}{2}~, \quad
{\bar{J}}_{\text{particle}}
= \frac{k}{r_++r_-} \frac{E+L}{2}~,\quad
T_{\text{particle}} = \bar{T}_{\text{particle}} = - \frac{km^2}{4}~. 
\end{align}

\subsection{The String Energy and Angular Momentum}
After embedding the geodesic solutions, we can apply the twist procedure:
\be
\phi \rightarrow \phi+w \sigma~,\qquad t \rightarrow t + w \tau~ \, ,
\ee
to obtain the wound solutions. 
At the level of the group element, this amounts to the transformation \cite{Hemming:2001we}:
\be
g \rightarrow e^{(r_+ -r_-)w x^+ \sigma^3/2} \, g \,  e^{-(r_+ + r_-)w x^-\sigma^3/2}~.
\ee
This in turn leads to the shifted currents
\begin{align}
J &= {J}_{\text{particle}} - (r_+-r_-) \frac{k w}{2}
~, \qquad \bar{J} = {\bar{J}}_{\text{particle}} - (r_++r_-) \frac{kw}{2} \, ,
\end{align}
and the shifted stress tensor for the solution in the twisted sector:
\begin{align}
\label{TTbartwisted}
T &=T_{\text{particle}} - w(r_+-r_-) J_{\text{particle}}+\frac{k}{4}(r_+-r_-)^2 w^2\\
\bar T &= \bar T_{\text{particle}} - w(r_++r_-) \bar J_{\text{particle}}+\frac{k}{4}(r_+ + r_-)^2 w^2~. \nonumber 
\end{align}
These shifts  implement the twisting operation in the sigma model \cite{Maldacena:2000hw, Hemming:2001we,Troost:2002wk}. 
Given these currents, it was observed in \cite{Natsuume:1996ij,Hemming:2001we} that the angular momentum generator $L_{\text {string}}$ encoded in the level matching condition is:
\begin{align}
L_{\text {string}} &=  
-(r_+-r_-) {J} + (r_++r_-) {\bar{J}} + \frac{k}{2} w J_{BH}~. \label{LorentzianQuantizationOfAngularMomentum}
\end{align}
{From} now on, we add an explicit  index to the black hole angular momentum $J_{BH}$ to distinguish it from the current component $J$.
 The string energy $E{\text {string}}$ equals: 
\begin{align}
E_{\text {string}} &=  
(r_+-r_-) {J} + (r_++r_-) {\bar{J}} ~.
\end{align}
In our subsequent discussions we reserve the symbol $E$ to indicate the particle energy and $E_{\text {string}}$ to denote the energy of a winding  string solution.

\subsection{The Energy of Wound Geodesics}
Presently, we analyze the winding strings  in more detail, following a similar analysis performed in the three-dimensional anti-de Sitter space-time \cite{Maldacena:2000hw}. 
We discuss two cases: a time like geodesic and a space like geodesic with large enough energy such that $E^2-L^2+m^2 M >0$.  We parameterize the solutions slightly differently from earlier work on the subject and we explicitly solve the on-shell conditions. Later on, we will link up these results with an asymptotic probe string analysis performed in section \ref{LongStringProbe} and  compute the Wess-Zumino-Witten orbifold choice for the integration constant $c$ in the two-form flux potential discussed in section \ref{Definition}.

\subsubsection{Winding Time-like Geodesics }
We first focus on the time-like case and assume that the initial particle energy $E$ as well as the mass parameter $m$ are positive. The on-shell conditions in conformal gauge on the string world sheet are
\begin{equation}
    T + h = 0~, \qquad \bar T + \bar h = 0~,
\end{equation}
where $(h,\bar h)$ are the conformal dimensions of operators in the internal sector of a consistent string background. The energy-momentum components $T$ and $\bar T$ are given in \eqref{TTbartwisted}, with the particle energy-momentum $T_{\text{particle}}$ computed in equation \eqref{JTparticle}. Solving the on-shell conditions for the currents and substituting them back into the expression for the string energy we find
\begin{equation}
\label{Estringoriginal}
E_{\text {string}} = \frac{1}{w}(h+\bar h-\frac{km^2}{2}) - \frac{kw}{2}(r_+^2 + r_-^2)~ \, .
\end{equation}
For time-like geodesics, since $m^2>0$ we  find the inequality
\be
E_{\text {string}} < \frac{1}{w}(h+\bar h) - \frac{kw}{2}(r_+^2 + r_-^2)~. \label{InequalityTimeLike}
\ee
It is useful to parameterize the particle energy $E$ as follows:
\be
E= m \widetilde E~.
\ee
We substitute this parameterization into the on-shell conditions and solve for the particle mass $m$ and angular momentum $L$ to find 
\be
\label{msoln}
m = - \widetilde E w \pm  \sqrt{\frac{2h+2\bar h}{k}+ w^2(\widetilde E^2+r_+^2 + r_-^2) }~, \quad L = \frac{\bar h - h}{kw} + r_+\, r_-\, w~.
\ee
We observe that if we choose the positive root for the mass $m$, then in the limit of large $\widetilde E$ the mass $m$ goes to zero. We can take this limit such that the particle energy $E$ remains finite. The string energy then takes the form
\be
E_{\text {string}} = \frac{h+\bar h}{w} - \frac{k w}{2}(r_+^2 + r_-^2) + O((\frac{m}{E})^2) ~.
\label{LargeRadiusEnergy}
\ee
At larger energy $E$, or, at fixed energy $E$ and smaller mass $m$, the time-like particle geodesic reaches 
further out in the black hole space-time. The same statement holds for the wound string. Thus, the limit we are considering is one in which the string gets closer to the boundary of space-time. We have computed the energy of the string in that limit.

Finally, in order to facilitate comparison with a later calculation, we consider a short twisted string at its turning point. For simplicity we set the angular momenta to zero: $J_{BH} =L=0$. From the geodesic solution \eqref{TLgeo}, we conclude that at the turning point we have a maximal radius
\be
\label{rmaxvsEtilde}
r_{\text{max}}^2 = \frac{E^2+M m^2}{m^2} = \widetilde{E}^2 + r_+^2 ~.  
\ee
This relates the particle energy and maximal radius of the short string. Our final goal is rather to  express the string energy $E_{\text {string}}$ in terms of the maximum radial extent $r_{\text {max}}$.
For this purpose, we concentrate on the setting with no internal excitation,  $\bar h= \bar h=0$. We then have  
\begin{align}
E_{\text {string}} &=  - \frac{k m^2}{2w} - \frac{k w r_+^2}{2}
\end{align}
Substituting the value \eqref{msoln} for $m$, we find the energy of the string in terms of the  ratio $\widetilde{E}$. 
Finally, rewriting the energy $\widetilde E$ in terms of $r_{\text {max}}$ using equation \eqref{rmaxvsEtilde},  we find
\begin{align}
\label{EstringWS}
 E_{\text {string}}  
&= -k\, w\, r_{\text {max}} \left(r_{\text {max}} - \sqrt{r_{\text {max}}^2-r_+^2}\right) \, .  
\end{align}
This is the relation between the energy $E_{\text {string}}$ of the short string and the maximal radius $r_{\text {max}}$ it attains at zero angular momentum. The energy of a twisted short string in the background of a black hole with angular momentum is computed in Appendix \ref{generalprobe}.

\subsubsection{Winding Space-like Geodesics}

We turn to the winding space-like geodesics for which $m^2$ is negative. It is convenient to define the parameter $\kappa$ through
\be
m = i\, \kappa~, 
\ee
and choose  $\kappa$ to be a positive real number. We solve the on-shell conditions  and obtain the expression for the  string energy in the twisted sector:
\be
E_{\text {string}} = \frac{1}{w}(h+\bar h + \frac{k\, \kappa^2}{2}) - \frac{kw}{2}(r_+^2 + r_-^2)~.
\ee
Since $\kappa^2 >0$ we obtain the inequality
\be
E_{\text {string}} > \frac{1}{w}(h+\bar h) - \frac{kw}{2}(r_+^2 + r_-^2)~.
\ee
This inequality is  complementary to the inequality (\ref{InequalityTimeLike}) for time-like excitations.
For the space-like case we parameterize the energy as 
\be
E = \kappa\, \widetilde{E}~,
\ee
and substitute into the on-shell conditions to solve for the imaginary mass $\kappa$ and the angular momentum $L$: 
\be
\kappa =  \widetilde E w \pm  \sqrt{ w^2(\widetilde E^2-r_+^2 - r_-^2)-\frac{2h+2\bar h}{k} }~, \quad L %
= \frac{\bar h - h}{kw} + r_+\, r_-\, w~.
\ee
We choose the negative root for the parameter $\kappa$ and observe as before that one can take the limits $\widetilde E\rightarrow \infty$ and $\kappa\rightarrow 0$ such that the energy $E$ remains finite. Substituting this into the expression for the string energy we obtain 
\be
E_{\text {string}} = \frac{h+\bar h}{w} - \frac{k w}{2}(r_+^2 + r_-^2) + O((\frac{\kappa}{E})^2) ~.
\ee
This is identical to the result obtained for the time-like geodesic. This suggests that there is a transition between the two solutions at $m^2=0=\kappa^2$, while keeping the energy finite. From the point of view of the time-like geodesic it is clear that the transition takes place at large radius. In this respect, the analysis is similar to the one in $AdS_3$ \cite{Maldacena:2000hw}. 

There is nevertheless an important difference between the analysis of winding strings in the BTZ black hole background as compared to  the winding strings in  $AdS_3$. In the latter case, it was observed that as the internal conformal dimension reached a critical value, given by $h=\bar h =kw^2/4$, the mass $m$ (or the parameter $\kappa$) tended to zero and there was a transition between the short string and long string solution for finite values of the rescaled energy $\widetilde E$ \cite{Maldacena:2000hw}.
The critical value of the internal conformal dimension $h$ at which  one obtains an exactly vanishing mass $m$ independently of the rescaled energy $\widetilde{E}$ would occur in the BTZ background at a negative, unphysical internal conformal dimension given by $h = \bar h = -kw(r_+^2 + r_-^2)/4 $. Thus, in the BTZ background, we see the short to long string transition only in the scaling limit.

\section{The Winding String Probing a Black Hole}
\label{LongStringProbe}
\label{NambuGoto}
In this section, we analyze the Nambu-Goto action for a winding string near the boundary of the three-dimensional black hole supplemented with NSNS flux.  This is a variation on the long string analysis in \cite{Seiberg:1999xz} in anti-de Sitter space-time. It will allow us to calculate the energy of the winding string more directly, and to understand which choice of NSNS two-form potential is  made in the Wess-Zumino-Witten model. Moreover, we will show how the choice of definition of the model shows up in   Fefferman-Graham coordinates.

\subsection{The Winding String Probe}
\label{BTZlong}
We study a winding string near infinity. The black hole metric asymptotes to the $AdS_3$ metric to leading order. However, it is known that the leading tension and volume contributions of a fundamental string in an asymptotically $AdS_3$ background with NSNS flux cancel. The naively subdominant terms become the leading non-zero terms and they depend on the BTZ black hole background under consideration. This is related to the fact that a fundamental string probe is a $1/2$ BPS object in  string theory and  has a charge equal to its mass.  

We consider a  string that winds $w$ times at large radius $r$.
We  plug this ansatz  into a Nambu-Goto action supplemented with a $B_{(2)}$  flux  term. We take a  $B_{(2)}$ field potential equal to -- see section \ref{Definition} --:
\begin{equation}
\label{B-fieldbkgd}
B_{(2)} = \frac{1}{2} (r^2-c) \, dt \wedge d \phi 
\, .
\end{equation}
For starters, we concentrate on the leading and subleading terms in a $1/r^2$ expansion. 
The BTZ metric in the $r^{-2}$ expansion reads:
\begin{align}
ds^2 &= \frac{dr^2}{r^2} + r^2 (-dt^2 + d \phi^2)
\nonumber \\
 & + \frac{dr^2}{r^4} (r_+^2 + r_-^2)+ (r_+^2+r_-^2) dt^2 -2 r_+ r_- dt d \phi  
 \nonumber \\
 & + O(r^{-6}) dr^2 \, .
\end{align}
All further corrections are proportional to  $dr^2$. 
The action $S$ is 
\begin{equation}
S = S_{NG} + S_B \, ,
\end{equation}
where the flux term in the action  is $S_B$ and $S_{NG}$ is the Nambu-Goto action. We choose the static gauge appropriate to describe a long string that winds $w$ times around the  circle:
\be
\tau=  t~,\qquad \phi=w \sigma~, 
\ee
and consider the radius $r(\tau, \sigma)$ to be an arbitrary function of the world sheet coordinates. We find the flux term:
\begin{equation}
S_B = \frac{l^2}{4 \pi \alpha'} \int_0^{2 \pi} d \sigma \int d \tau \epsilon^{\alpha \beta} \partial_\alpha X^\mu \partial_\beta X^\nu B_{\mu \nu} =  \frac{kw}{2 \pi} 
\int d \tau  d \sigma ~ (
r(\tau, \sigma)^2
-c)
\, .
\end{equation}
The Nambu-Goto action $S_{NG}$ is equal to:
\begin{align}
S_{NG} &= -\frac{l^2}{2 \pi \alpha'} \int d \tau \int_0^{2 \pi} d \sigma  \sqrt{ -\det \partial_\alpha X^\mu \partial_\beta X^\nu G_{\mu \nu}} \, .
\end{align}
The argument of the square root is the determinant of the induced metric $h_{\alpha\beta}$ on the worldvolume of the long string. The components of the induced metric have the following leading terms in the large $r$ expansion:
\begin{align}
h_{\tau\tau} & =-r^2 + r_+^2 + r_-^2  + \left(\frac{\dot r}{r}\right)^2 ~, \quad
h_{\tau\sigma} = -r_+r_- w+ \frac{\dot r r'}{r^2}~, \quad 
h_{\sigma\sigma} = w^2 r^2 + \left(\frac{ r'}{r}\right)^2~.
\end{align}
 In the total action including the flux term, the leading volume and area contributions cancel and using $l^2 = k\, \alpha'$, we obtain
 \begin{align}
  S  = \frac{k w}{4 \pi} \int d \tau d \sigma \, \left( r_+^2 + r_-^2-2c) + \frac{1}{r^2}\left(\dot r^2 - \frac{1}{w^2}(r')^2 + \frac{1}{4}(r_+^2 - r_-^2)^2 \right) \right) + \ldots
 \end{align}
 We recognize a non-trivial kinetic term and a leading potential term that gives the string an effective negative potential energy:
 \begin{equation}
 V =  -\frac{kw}{2} (r_+^2+ r_-^2-2c)  \, .
 \end{equation} 
 The final result  depends on the choice of constant $c$. 
If we identify the potential energy with the energy (\ref{LargeRadiusEnergy}) computed in the Wess-Zumino-Witten model for an on-shell large radius string, we find that we must set the constant $c$ to zero. 
 If we do set the constant $c$ to zero, then the $AdS_3$ result  $V= kw/2$ \cite{Maldacena:2000hw, Seiberg:1999xz} is recovered by analytic continuation in the outer horizon radius: $r_+^2=M \longrightarrow -1$ (which indeed continues the BTZ metric \eqref{BTZMetric} into the global $AdS_3$ metric).\footnote{In the anti-de Sitter space-time, it  can be  straightforwardly argued that $c=0$ is a  necessary condition for regularity in the interior.} Therefore our result is
 consistent with the result of \cite{Maldacena:2000hw, Seiberg:1999xz}. Our calculation extends the result to the BTZ black hole backgrounds and provides information on the implicit choice of flux term in the orbifold Wess-Zumino-Witten model. 

\subsubsection{The Energy of the Probe String}
 
Finally, we perform a check of the fact that the integration constant is zero  in the Wess-Zumino-Witten model to all orders in the radius expansion, by calculating the exact Nambu-Goto  string energy  and comparing with the result obtained using the world sheet Wess-Zumino-Witten model. For this purpose let us compute the energy  near the turning point in the simple case of $r_-=0$. We set the derivatives to zero, $\dot r = 0= r' $ and introduce $r=r_{\text {max}}$, the maximum radius attained by the string. The string energy is then  given by the (negative) integral over  the Lagrangian:
\begin{align}
E_{\text {string}} &=  - \int d\sigma~ L 
 = -k w \left( (r_{\text {max}}^2 - c) - r_{\text {max}} \sqrt{r_{\text {max}}^2-r_+^2} \right) \, . 
\end{align}
Only for zero integration constant $c$ the  string energy matches the Wess-Zumino-Witten result \eqref{EstringWS}.The equality of the Nambu-Goto and Wess-Zumino-Witten energies in the presence of a non-zero black hole angular momentum $J_{BH}$ is proven in Appendix \ref{generalprobe}. 
 
\subsubsection{The Winding String as a Probe in Three-dimensional Gravity}
One may wonder how the effective tension of the string changes due to the presence of the black hole. As a first step in an understanding, we note that in  three-dimensional gravity, a point particle creates a deficit angle. A second static point particle far removed sees a trivial metric and does not feel a force due to the presence of the first particle. However, a non-local excitation like a winding string will notice that the asymptotic circle has shrunk. We can either consider that the string winds a smaller circle, or that its effective tension has shrunk. Our black hole set-up is different, but the physical energy of the winding asymptotic string still depends on the mass $M$ of the black hole in the interior. That is a long range three-dimensional gravity effect of the black hole mass on the non-local  winding probe. 

Finally, note that the effective tension of the winding string is zero for a massless BTZ black hole.  One way to understand this is via the interpolation between the $AdS_3$ space and the massless  black hole. The conical deficit angle gradually eats up the whole space.  Another way to understand the vanishing effective tension is to recall that the massless BTZ black hole is like three-dimensional anti-de Sitter space in the Poincar\'e patch, with a cylinder boundary. The zero result we find for this case is consistent with the result obtained in \cite{Seiberg:1999xz}.\footnote{The Liouville gap in conformal dimension that is invoked on the sphere in \cite{Seiberg:1999xz} is semi-classically absent on the cylinder.}

\subsection{The Winding String  in the Fefferman-Graham Expansion}
\label{FGlong}
Our analysis raises the question of how the asymptotic analysis of  \cite{Seiberg:1999xz} in Fefferman-Graham coordinates  accommodates  these potential terms. 
In this section, we make the link between the determination of the long string action in Fefferman-Graham coordinates, and our analysis in BTZ coordinates. The link relates to the remarks made in section \ref{Definition}.

The starting point for the analysis  \cite{Seiberg:1999xz} is the asymptotically $AdS_3$ metric in Fefferman-Graham coordinates:\footnote{Note that the  black hole metric (\ref{BTZMetric}) in BTZ coordinates is of a different form.}
 \begin{equation}
 ds^2 = l^2 ({d\xi^2} + \frac{e^{2\xi}}{4} g_{ij} dx^i dx^j
 -P_{ij} dx^i dx^j + \dots) \, .
 \end{equation}
 where  the tensor $P$ has a trace determined by the curvature $R$ of the leading boundary metric $g$
 \begin{equation}
 g^{ij} P_{ij} = \frac{R}{2} \, 
 \end{equation}
 and is otherwise undetermined by the bulk equations of motion. The tensor $P$ is closely related to the boundary energy-momentum tensor. 
 The Fefferman-Graham expansion is exact in two dimensions, and only needs one further subleading correction. 
 
When one works in the Euclidean and assumes that one can cap off the asymptotically $AdS$ space in the interior, then the result of \cite{Seiberg:1999xz} on the action of a long string is universal in the sense that it is independent of the space-time boundary energy-momentum tensor. The result for the long string action is then \cite{Seiberg:1999xz}:
\begin{align}
 S 
&= w \, l^2 \int d^2 x  \sqrt{-g}  (\frac{R}{4} - \xi \frac{R}{2} 
 - \frac{1}{2} g^{\alpha \beta} \partial_\alpha \xi \partial_\beta \xi 
+ O(e^{-2\xi}))
 \, .
 \end{align}
Compared to \cite{Seiberg:1999xz}, we have added an overall factor of $w$ for a string that winds the angular direction $w$ times. 
Here, we also consider Lorentzian space-times in which there is not necessarily a unique way to cap off  the interior (as in the Lorentzian BTZ black hole). In section \ref{Definition}, we have discussed the resulting  ambiguity in the long string action. It comes down to adding a constant contribution to the flux term in the action. The resulting final action is:
 \begin{align}
 S 
&= w \, l^2 \int d^2 x  \sqrt{-g}  (\frac{R}{4} - \xi \frac{R}{2} 
 - \frac{1}{2} g^{\alpha \beta}  \partial_\alpha \xi \partial_\beta \xi  
+\tilde{c}+ O(e^{-2\xi}))
 \, .   \label{FGaction}
 \end{align}
The action for the scalar $\xi$ on the long string world sheet contains a linear dilaton  deftly exploited in \cite{Seiberg:1999xz}. The linear dilaton slope is set by the cosmological constant.

 \subsubsection*{Discussion }

 To compare the approaches of subsections \ref{BTZlong} and \ref{FGlong}, it is useful to  relate the BTZ and Fefferman-Graham coordinate systems. The BTZ metric in Fefferman-Graham coordinates is found through the coordinate transformation $e^{\xi}/2=\rho(r)$ that satisfies:
 \be
 \frac{r^2\, dr^2}{(r^2-r_+^2)(r^2-r_-^2)} =\frac{d\rho^2}{\rho^2}~.
 \ee
 Solving for the radial coordinate $\rho(r)$ we find the  solution in the regime $r>r_+ $ beyond the outer horizon:
 \be
 \rho(r) = \sqrt{r^2-r_+^2} + \sqrt{r^2 - r_-^2}~.
 \ee
 %
We also
define the null coordinates $w^\pm$:
\be
w^{\pm} = \frac12(\phi \pm t)~.
\ee
The metric in these  coordinates is 
\begin{align}
ds^2 
&= \frac{d\rho^2}{\rho^2}+ \rho^2\left(dw_+ + \frac{(r_+ + r_-)^2}{\rho^2} dw_-\right) \left(dw_- + \frac{(r_+ - r_-)^2}{\rho^2} dw_+ \right) \, .
\end{align}
We compare the flux terms in the respective coordinate systems. They both start out with a quadratic divergence at large radius. At large radius the coordinate systems are related by a  shift in the radius squared. This shifts shows up as a constant term in the flux contribution to the action. More in detail, the transformation of the flux term is:
\begin{equation}
 \int (r^2-c) \, d \phi dt = \int (\frac{\rho^2}{4} + \frac{r_+^2+r_-^2}{2} - c + \dots) \, d\phi dt \, .
\end{equation}
Thus, the  constant flux term $c$ in the BTZ coordinates and  the constant $\tilde{c}$ in the Fefferman-Graham coordinates are related as:
\begin{equation}
c = \tilde{c} - \frac{r_+^2+r_-^2}{2} \, .
\end{equation}
For instance, if $c=0$, then $\tilde{c} = (r_+^2+r_-^2)/2$ and the potential term in the action (\ref{FGaction}) agrees with the potential energy  found in section \ref{Classical}.
The choice of constant is reflected here in a choice of parameterization of the radial coordinate at infinity or more precisely  in a choice of radial cut-off as discussed in section \ref{Definition}. This clarifies the interplay between the asymptotic long string analysis in Fefferman-Graham coordinates \cite{Seiberg:1999xz}, in BTZ coordinates, as well as the flux term in the action.

\section{The Partition Function and the Euclidean Twist}
\label{Euclidean}
\label{PartitionFunction}
\label{CosetOrbifold}
In this section, we analyze properties of string theory in   Euclidean BTZ black holes in three-dimensional anti-de Sitter space-time. We highlight the differences between the one loop vacuum amplitude in the black hole background and the thermal $AdS_3$ vacuum free energy. We compute the  world sheet conformal field theory partition function on the BTZ background. It  can be interpreted as the  partition function corresponding to the $\mathbb{Z}$ orbifold that gives rise to the black hole.  We describe the twisted sectors of the conformal field theory and how they fit into the classical theory. We comment on how to fix the flux term in the Euclidean theory. 
We refer to 
\cite{Mertens:2014nca,Mertens:2015ola} for interesting and closely related results.  We believe our analysis further clarifies the relation between the spectrum and the partition function. We stress   the important differences that appear in the analytic continuation of the Euclidean results to their Lorentzian signature counterparts between the case of the anti-de Sitter and the black hole space-times. 

\subsection{The Interpretation of the Partition Function}
To interpret the black hole partition function, it is useful to contrast it with the one loop, finite temperature free energy of string theory in three-dimensional anti-de Sitter space. The latter is obtained by compactifying the  $AdS_3$ space-time on a thermal circle \cite{Polchinski:1985zf,Maldacena:2000kv}. The resulting partition function depends on two parameters (apart from the cosmological constant), which are the inverse temperature $\beta$, and the chemical potential for angular momentum $\mu \beta$. We thus have a multi-string free energy, related to the one-loop vacuum energy $Z_{EAdS_3}(\beta,\mu \beta)$ \cite{Polchinski:1985zf,Maldacena:2000kv}. It is obtained by thermal compactification, which is implemented through the trace:
\begin{equation}
Z_{EAdS_3} (\beta,\mu \beta) = Tr_{{\cal H}} e^{-\beta H + i \mu \beta L} \, .
\end{equation}
The Hamiltonian $H$ implements time translation, the operator $L$ corresponds to the rotation generator and we trace over the Hilbert space ${\cal H}$ to make this a one-loop vacuum amplitude. To lay bare the relation between the thermal amplitude and the Lorentzian spectral interpretation requires further extensive analysis \cite{Polchinski:1985zf,Maldacena:2000kv,Ashok:2020dnc}. 

The nature of the BTZ one-loop vacuum amplitude is different. It is a function of two parameters, both specified by the Euclidean black hole background itself. They are the mass and angular momentum of the black hole or the inner and outer horizon radii. The partition function takes the form of an orbifold partition function with twisted sectors labelled by the winding number $w$ and a projection operator that guarantees invariance under the orbifold group:
\begin{equation}
Z_{BTZ (M,J)} = \sum_w Tr_{{\cal H}_w} e^{2 \pi i w L} \, .
\end{equation}
The Hilbert space splits up into twisted sectors labelled by a winding number $w$.  The orbifold projection operator that identifies the geometry is related to the angular translation generator $L$ in the Euclidean coset conformal field theory on  $H_3^+=SL(2,\mathbb{C})/SU(2)$.

 We conclude that we are performing a trace in which we glue by a complexified angular momentum (instead of gluing by a complexified energy). While for anti-de Sitter space we have a partition function with two parameters, for each BTZ background we have a single partition function. In the first case, we can couple two parameters to two charges. In the second case, we can analyze the trace as a function of the (two parameter) background.  
The BTZ Hilbert space depends on the radii $r_\pm$ and it is this dependence that is measured by the trace.

We refer to  Appendix \ref{Geometries} for a more elaborate discussion of the geometry, the angular Killing vector and the  identification
 in the Euclidean model.

  \subsection{The Sigma Model on the Euclidean Black Hole}

It is useful to specify a few expectations we have for the orbifold spectrum, directly in the Euclidean context.  The Euclidean coset model is often presented in  Poincar\'e coordinates $(\xi,\gamma,\bar{\gamma})$:
\begin{align}
ds^2 &= d \xi^2 + e^{2 \xi} d \gamma d \bar{\gamma}
\end{align}
with the two-form potential $B_{(2)}$ 
\be
B_{(2)} = \frac{1}{2} e^{2\xi}\, d\gamma\wedge d\bar\gamma~.
\ee
The $B$-field is imaginary and the worldsheet theory is not reflection positive. We will be concerned with Euclidean worldsheets for which the $B$-field contribution to the worldsheet action is real \cite{Giveon:1998ns}: 
\begin{align}
    S= \frac{k_b}{\pi} \int d^2z~(\pa\xi\bar\pa\xi+ e^{2\xi}\bar \pa\gamma \pa\bar\gamma )~.
\end{align}
We have defined the  complex world sheet coordinates $z =i\tau_E + \sigma$, and $\bar z= -i\tau_E + \sigma$.
In the Appendix we show how to relate the $(\xi,\gamma,\bar\gamma)$ coordinates to the  Wick rotated BTZ coordinates $(r, t_E, \phi)$.
Using the explicit change of variables in \eqref{Poincaretortphi}, one can check that the $B_{(2)}$ -field is strictly gauge equivalent to:
\be
B_{(2)} = \frac{i}{2} \, (r^2- r_+^2) dt_E\wedge d\phi~.
\ee
The integration constant $c$, which was discussed in detail in our discussion of the Lorentzian theory, now takes the non-zero value $c=r_+^2$.
In the Euclidean, the time circle pinches off at the outer horizon and the demand to avoid a source term at the tip fixes the value of the integration constant.\footnote{The demand that the three-form flux be regular at the tip is not necessarily compatible with other reasonable demands. See e.g. \cite{Ferrari:2016vcl} for a discussion in the context of matching probe and gravity free energies.}

 \subsection{Expectations for Twisted Sectors}

The classical coset sigma-model  has solutions to the equations of motion that follow from the factorized solutions for the model on $SL(2,\mathbb{C})$ (see e.g. \cite{deBoer:1998gyt}):
\begin{align}
\xi(z,\bar{z}) &= \log(1+b(z) \bar{b} (\bar{z})) + \rho(z) + \bar{\rho}(\bar{z})
\nonumber \\
\gamma(z,\bar{z}) &= a(z) + \frac{e^{-2 \rho(z)} \bar{b}(\bar{z}) }{1+b(z) \bar{b}(\bar{z})}
\nonumber \\
\bar{\gamma}(z,\bar{z}) &=  \bar{a}(\bar{z}) + \frac{e^{-2 \bar{\rho}(\bar{z})} {b}(z) }{1+b(z) \bar{b}(\bar{z})}
\, ,
\label{EuclideanClassicalSolutions}
\end{align}
where the functions on the right hand side are holomorphic or anti-holomorphic. 

\subsubsection{The Twisted Euclidean Classical Solutions}

As we saw in Section \ref{windstringssec}, given a Lorentzian classical string solution to the equations of motion, it is possible to find a winding string solution by performing a twist. In the Euclidean, there is an analogous operation that maps one solution into another: 
\be
\label{twistontausigma}
t_E \rightarrow t_E + w \tau_E ~,\qquad \phi \rightarrow \phi + w \sigma~.
\ee
In the Appendix \ref{BTZmetriccoordinates}, we relate the BTZ coordinates to  coordinates $(\varphi, v, \bar v)$ suitable for world sheet path integration \cite{Gawedzki:1990ji}. See equation \eqref{identifone}. Those  coordinates we can in turn relate to the Euclidean Poincar\'e coordinates:
\be
\label{vvbargammagammabar}
\gamma = e^{\phi}\,v~,\quad \bar \gamma = e^{\phi}\, \bar v~, \quad  \xi= -\varphi~.
\ee
The Euclidean twist action  \eqref{twistontausigma} then can be translated to the $(\varphi, v,\bar v)$ variables, 
and subsequently to the $(\xi, \gamma,\bar\gamma)$ coordinates: 
\begin{align}
\gamma &\longrightarrow e^{ w(r_+ + i r_-) z}\, \gamma~, \quad \bar\gamma \longrightarrow e^{ w(r_+-i r_-)\bar z}\, \bar\gamma~, \nonumber \\
\xi &\longrightarrow \xi-\frac{ w}{2}\left((r_+ + ir_-) z + (r_+ - i r_-) \bar z\right)~.
\end{align}
Comparing with the general classical solutions \eqref{EuclideanClassicalSolutions}, we find that the twisted solutions are obtained from the untwisted ones by the action: 
\begin{align}
a(z)  &\longrightarrow  e^{i w (r_--i r_+) z}~a(z)~,\quad 
 b(z)  \longrightarrow b(z)~,\quad 
\rho(z) \longrightarrow \rho(z) - i\frac{w}{2}(r_--ir_-) z ~ ,
\end{align}
and analogously for the complex conjugate functions $\bar{a}$, $\bar{b}$ and $\bar{\rho}$.

\subsubsection{The Euclidean Coset Currents and Their Twists}

As in the Lorentzian model, we wish to understand the action of twisting on the classical currents and energy momentum tensors. 
The $sl(2,\mathbb{C)} $ current generators in the  $(\xi,\gamma,\bar{\gamma})$ coordinates are (see e.g. \cite{Kutasov:1999xu}):
\begin{align}
J^- &= i\, k e^{2 \xi} \partial \bar{\gamma}
\nonumber \\
J^3 &= -i\, k(\partial \xi  - \gamma e^{2 \xi} \partial \bar{\gamma})
\nonumber \\
J^+ &=-i\, k (\partial \gamma +2 \gamma \partial \xi - \gamma^2 e^{2 \xi} \partial \bar{\gamma}) \, .
\end{align} 
Thus, the action of twisting on the Euclidean current generators is:
\begin{align}
J^- &\rightarrow e^{-  w (r_+ +ir_-) {z}} J^-
\nonumber \\
J^3  & \rightarrow J^3 +i \frac{kw}{2} (r_+ + i r_-)
\nonumber \\
J^+ & \rightarrow  e^{ w (r_+ -ir_-) {z}} J^+  \, .
\end{align}
For the barred $sl(2,\mathbb{C})$ currents, similar twisting rules hold. The energy-momentum tensor component in terms of these currents is 
\begin{align}
T &= - \frac{1}{2k}(J^- J^+ + J^+ J^-)+ \frac{1}{k} (J^3)^2
\nonumber \\
&=-k((\pa\xi)^2 + e^{2\xi} \pa \gamma \pa\bar\gamma)~.
\end{align}
The anti-holomorphic stress tensor can be similarly written with the substitution $\partial \rightarrow \bar{\partial}$.
The twisting action on the Euclidean energy momentum tensor is:
\begin{align}
\label{twistedT}
T & \longrightarrow T +  w (r_+ + i r_-)(i J^3)  + \frac{k w^2}{4} (r_+ + i r_-)^2  \, , \\
\bar{T} & \longrightarrow \bar{T} +  w (r_+-i r_-)(-i \bar{J}^3)   + \frac{k w^2}{4} (r_+ - i r_-)^2 \, .
\end{align}
 We observe that the Lorentzian twists \eqref{TTbartwisted} are obtained by analytic continuation of the inner horizon radius $r_-$.
By the same analytic continuation procedure, or by analysing level matching directly in the Euclidean, one can write down  the Euclidean angular momentum generator: 
\begin{align}
L_{\text {string}}^E   
&=  
-(i J^3)(r_+ + i r_-) +(-i\bar J^3)(r_+ - i r_-) +kw (-ir_- r_+)  \, .
\label{AngMomEucl}
\end{align}
We have analyzed classical properties of the Euclidean twisted sectors. These classical properties will be reflected in the world sheet conformal field theory partition function.

\subsection{The Partition Function}

 In this section we calculate the one loop string amplitude in the Euclidean black hole background and in particular the  BTZ world sheet partition function factor. Furthermore we speculate on its Lorentzian interpretation.
 
 The partition function is a weighted sum over the quantum spectrum. It can be computed by adapting the calculation of the thermal $AdS_3$ partition function \cite{Maldacena:2000kv}. For the purpose of computing the partition function, it is most convenient to write the worldsheet action for strings propagating in the coset orbifold model in the $(\varphi, v, \bar v)$ coordinates -- we refer to Appendix \ref{Geometries} and \cite{Maldacena:2000kv,Ashok:2020dnc} for the details --:
\begin{align}
    S= \frac{k_b}{\pi} \int d^2z~(\pa\varphi\bar\pa\varphi+(\pa\bar v + \bar v \pa\varphi)(\bar\pa v+v \bar \pa\varphi))~,
\end{align}
where we have denoted the quantum level of the bosonic $sl(2,\mathbb{R})$ current algebra by $k_b$. As shown in the Appendix, the coordinates undergo a non-trivial identification in the black hole orbifold -- see equation \eqref{identiftwo}:
\be\label{phivvbarident}
(\varphi, v,  \bar v) \sim (\varphi + 2\pi r_+, e^{2\pi i r_-} v, e^{-2\pi i r_-} \bar v)~,
\ee
where $r_{\pm}$ are the inner and outer horizon radii.
The one loop string amplitude on the background is
\be
Z =  \int_{{\cal F}_0} \frac{d^2\tau}{4 \tau_2^2} Z_{BTZ}(r_{\pm})~Z_{gh}~Z_{int}~,
\ee
where the ghost and internal sector partition functions are given by
\begin{align}
Z_{\text{gh}} &=\tau_2~ |\eta(\tau)|^4~,\\
\Zint &= (q\bar q)^{-\frac{\cint}{24}}\sum_h d(h,\bar h)~q^h {\bar q}^{\bar h}~,
\end{align}
$q=e^{2 \pi i \tau}$ is the elliptic nome and $d(h,\bar{h})$ denotes a degeneracy of the states of the internal conformal field theory.
Given the coordinate identifications  \eqref{phivvbarident} and comparing with the thermal $AdS_3$ calculation, one can  write down the Euclidean conformal field theory partition function by a suitable replacement of the fugacities $(\beta,\mu \beta)$ with the background horizon radii $(r_+,r_-)$ -- see Appendix \ref{DeterminationPartitionFunction} --:
\begin{equation}
Z_{BTZ} = \frac{r_+ \sqrt{k_b-2}}{ \sqrt{\tau_2}}
\sum_{w,m} \frac{e^{-\pi \frac{k_b}{ \tau_2} r_+^2 |m-w \tau|^2 + \frac{2 \pi}{\tau_2} \Imag (\bar{U}_{m,w})^2}}{|\theta_1(\bar{U}_{m,w},\tau)|^2} \, .  \label{StartingPoint}
\end{equation}
We have defined the holonomy:
\begin{equation}
\bar{U}_{m,w} = 
(r_- - ir_+) (m-w\tau) \,. \label{BTZHolonomy}
\end{equation}
Note that the holonomy $\bar{U}$ transforms as $\bar{U} \rightarrow \bar{U}/(c \tau+d)$ under world sheet modular transformations $\tau \rightarrow (a \tau + d)/(c \tau+d)$ which renders the partition function modular invariant.\footnote{ Either from this transformation property or from the definition of Ray-Singer torsion \cite{Ray:1973sb} -- see also \cite{Gawedzki:1990ji} -- it is clear that we must have the $\tau$ dependent holonomy $\bar{U}=\nu$ for the partition function factor $\theta_1(\nu,\tau)$. This corrects an inaccuracy in \cite{Maldacena:2000kv, Ashok:2020dnc} which is of no further consequence there. }
%
\subsubsection{An Analysis of the Spectrum}
In the remainder of this section we will develop arguments on how to interpret the Euclidean and Lorentzian spectral content of the partition function. Our arguments and analysis are speculative -- we hope they will be good guides for a more rigorous analysis. See also \cite{Mertens:2014nca,Mertens:2015ola} for a related but different discussion.

We expand the $\theta_1$-function in the denominator of the partition function \eqref{StartingPoint} in a series expansion using $z=e^{2\pi i \nu}$:
\be
\frac{1}{\theta_1(\nu,\tau)} = i \frac{1}{\eta^3}\sum_{r\in\mathbb{Z}} z^{r+\frac12} S_r(q)~,\quad\text{with}\quad S_r(q) = \sum_{n\ge 0}(-1)^n q^{\frac{n}{2}(n+2r+1) }~.
\ee
This expansion is valid only when $|q| < z < 1$. We ignore this substantial difficulty here and develop a further formal argument (similar in spirit to Appendix B of \cite{Maldacena:2000hw}).
The $\nu$ variable is given by the holonomy $\bar{U}_{m,w}$. As a result of our formal expansion, we miss out on certain contributions to the partition function. We will suggest an interpretation of some of these contributions in terms of continuous representations later on. At the moment we concentrate on  contributions of the twisted discrete states  to the partition function. Keeping this caveat in mind, we continue with the series expansion. 
We obtain  
\begin{align}
Z_{BTZ} &= \frac{r_+ \sqrt{k_b-2}}{  \sqrt{\tau_2}}\frac{1}{|\eta|^6} 
\sum_{w,m,r,\bar r} S_r S_{\bar r} ~ e^{2\pi i m r_-(r-\bar r)} ~ e^{2\pi m r_+(1+r+\bar r)}   \cr
& q^{-( r+\frac12)w(r_- -i r_+)}~ 
\bar{q}^{-(\bar r+\frac12) w(r_- + i r_+)}~ 
e^{-\pi \frac{k_b}{ \tau_2} r_+^2 |m-w \tau|^2 + \frac{2 \pi}{\tau_2} \Imag (\bar{U}_{m,w})^2}  \, .
\end{align}
We  use an integral representation that introduces a formal radial momentum $s$: 
\begin{equation}
\int_{-\infty}^{+\infty} ds~e^{-4 \pi \tau_2 \frac{s^2}{k_b-2}}
e^{-4 \pi i s \Imag(\bar{U}_{m,w})} = \frac{1}{2} \frac{\sqrt{k_b-2}}{\sqrt{\tau_2}} e^{-\frac{ (k_{b}-2) \pi \Imag(\bar{U}_{m,w})^2}{\tau_2}} \, .
\end{equation}
We then obtain
\be
\begin{aligned}
Z_{BTZ} &=  \frac{2r_+}{|\eta|^6} 
 \sum_{w,m,r,\bar r} S_r S_{\bar r} ~ e^{2\pi i m r_-(r-\bar r)} ~ e^{2\pi m r_+(1+r+\bar r)}  q^{-( r+\frac12)w(r_- -i r_+)}~ 
\bar{q}^{-(\bar r+\frac12) w(r_- + i r_+)} \cr
& ~ 
e^{-\pi \frac{k_b}{ \tau_2} r_+^2 |m-w \tau|^2 }~e^{\frac{ k_{b} \pi \Imag(\bar{U}_{m,w})^2}{\tau_2}}  \int_{-\infty}^{+\infty} ds~e^{-4 \pi \tau_2 \frac{s^2}{k_b-2}}
e^{-4 \pi i s \Imag(\bar{U}_{m,w})}
\end{aligned}
\ee
After simplification, we find 
\begin{align}
Z_{BTZ} &=   \frac{2r_+}{ |\eta|^6} 
 \sum_{w,m,r,\bar r} S_r S_{\bar r} ~ e^{2\pi i m r_-(r-\bar r)} ~ e^{2\pi m r_+(1+r+\bar r)}  q^{-( r+\frac12)w(r_- -i r_+)}~ 
\bar{q}^{-(\bar r+\frac12) w(r_- + i r_+)} \cr
& ~ e^{2\pi k_b m w r_+ r_-}~e^{-\pi k_b w^2( (r_+^2 - r_-^2)\tau_2 +2 r_+ r_- \tau_1)}
~\int_{-\infty}^{+\infty} ds~e^{-4 \pi \tau_2 \frac{s^2}{k_b-2}}
e^{-4 \pi i s(r_+(w\tau_1 -m) - r_- w\tau_2 )} 
\end{align}
We  observe that in the exponent, only the linear term in the wrapping number $m$ remains. Formally, the coefficient of the linear term in $m$ in the exponent needs to be an integer times $2 \pi i$ in order to pass Dirac's comb and we find the constraint:
\begin{equation}
 r(r_- - i r_+)- \bar r(r_- + i r_+)  - i k_b w r_- r_+ -2ir_+ (\frac12+is) \in \mathbb{Z}  \, .
\end{equation}
Defining a Casimir variable $j = \frac12 + i s$, we can rewrite the constraint as 
\begin{equation}
(j+ r)(r_- - i r_+) -(j +\bar r)(r_- + i r_+) - i k_b w r_- r_+  \in \mathbb{Z}  \, .
\label{EuclideanQuantizationOfAngularMomentum}
\end{equation}
We think of this quantization equation as indicating poles in the complex $j$ plane where discrete contributions to orbifold invariant correlation functions reside. We observe that this is precisely the condition  obtained by  analytic continuation of the Lorentzian quantization condition \eqref{LorentzianQuantizationOfAngularMomentum}, provided we identify 
\be
\label{J3J3bar}
J^3 = j+r~,\quad \bar J^3 = j+\bar r~. 
\ee
The partition function becomes
\begin{align}
Z_{BTZ}=- \frac{2ir_+}{|\eta|^6} 
\sum_{w,r,\bar r}  S_r S_{\bar r} ~\int_{-i\infty}^{+i\infty}~ dj~
& q^{- \frac{(j-\frac12)^2}{k_b-2}- (j + r)(r_-- i r_+)w -\frac{k_bw^2}{4}(r_- - i r_+)^2   } \cr
&\bar{q}^{- \frac{(j-\frac12)^2}{k_b-2} - (j+\bar r)(r_- +i r_+)w  -\frac{k_bw^2}{4}(r_- + i r_+)^2  } \, .
\end{align}
Finally, we  combine this result with the internal and ghost contributions to write down the one loop string amplitude. 
Firstly we rewrite the contribution of the internal conformal field theory and the descendants:
\begin{align}
 \frac{ \Zint}{|\eta(\tau)|^2}~ S_r~ S_{\bar r} 
 &=(e^{4\pi\tau_2})^{(1-\frac{1}{4(k_b-2)}) }\sum_{N, h,\bar N, \bar h}d_{r,h,N}~ q^{h+N}~ q^{\bar h+\bar N}~.
\end{align}
Lastly, we rotate the $j$ integration contour to coincide with the real line (up to an implicit Feynman type regularization). 
This allows us to write down the one loop string amplitude in the form
\begin{align}
Z_{BTZ} &= 
2\pi r_+ \int_{{\cal F}_0} \frac{d^2\tau}{\tau_2}  \sum_{w,r, h,N} d_{r,h,N} 
\int_{-\infty}^{+\infty} \frac{dj}{\pi}  ~ q^{L_0-1} \bar q^{{\bar L}_0 -1}~,
\end{align}
where the worldsheet Virasoro generators $L_0$ and $\bar L_0$ are given by
\begin{align}
L_0 &= -\frac{j(j-1)}{k_b-2} +i (j + r)(r_+ + i r_-)w +\frac{k_bw^2}{4}(r_+ + i r_-)^2 + h + N~,\\
\bar L_0&=  -\frac{j(j-1)}{k_b-2} -i (j + \bar r)(r_+ - i r_-)w +\frac{k_bw^2}{4}(r_+ - i r_-)^2 + \bar h + \bar N~.
\end{align}
We note that this precisely matches with the Virasoro generators \eqref{twistedT}  for  twisted short strings, provided we make the  identification of charges  \eqref{J3J3bar}.  After continuation to the Lorentzian, the speculation is that these discrete representation excitations are the quantum counterpart to the classical short winding strings based on time-like geodesics.

\subsubsection{Continuous Contributions} 
The $\theta_1$ function in the denominator in the partition function has a pole at holonomy $U_{m,w}=0$ among others. Expanding the $\theta_1$ function in disregard of the range of validity of the expansion is formally similar to neglecting such pole contributions \cite{Maldacena:2000hw}. In this subsection, we wish to show that the pole at $U_{m,w}=0$ allows for an interpretation in terms of characters of continuous representations. Inspiration for this identification is drawn from \cite{Maldacena:2000hw}.

We start once again from the initial partition function and make a sketch of the pole contribution:
\begin{equation}
Z^{div}_{BTZ}= \frac{r_+ \sqrt{k_b-2}}{ \sqrt{\tau_2}} \sum_{w,m} 
\frac{1}{|\eta(\tau)|^6} 
\delta(\bar{U}_{m,w}) \delta( U_{m,w})
\end{equation}
What we shall  do now is to rewrite this modular invariant,  divergent contribution in terms of the continuous characters. For this purpose we insert particular prefactors that vanish at the locations of the poles:
\begin{equation}
Z^{div}_{BTZ} = \frac{r_+ \sqrt{k_b-2}}{ \sqrt{\tau_2}} \sum_{w,m} 
e^{-\frac{\pi k_b r_+^2}{\tau_2} |m-w \tau|^2}
e^{ \frac{\pi k_b}{\tau_2} (\Imag U_{m,w})^2 }
\frac{1}{|\eta(\tau)|^6} 
\delta(\bar{U}_{m,w}) \delta( U_{m,w}) \, .
\end{equation}
We wish to write this delta-function contribution as an integral over twisted continuous characters. To that end we insert the integral representation:
\begin{align}
Z^{div}_{BTZ} &=2r_+
\int_{-\infty}^{+\infty}ds~ e^{-4 \pi \tau_2 \frac{s^2}{k_b-2}  
} 
\nonumber \\
& 
\sum_{w,m} 
e^{-\frac{\pi k_b r_+^2}{\tau_2} |m-w \tau|^2}
e^{ \frac{\pi k_b}{\tau_2} (\Imag U_{m,w})^2 }
\frac{1}{|\eta(\tau)|^6} 
\delta(\bar{U}_{m,w}) \delta( U_{m,w})
\, .
\end{align}
The discrete summation variable $m$ is Poisson dual to the quantized angular momentum. To obtain a quantized angular momentum in the continuous representations, the corresponding left and right generator eigenvalues need to be locked yet allow for an intercept that is tuned between the left and the right sector. To introduce the intercept and decouple the left and right quantum numbers, we introduce the integral:
\be
\int_0^1 ~d\alpha e^{2\pi i\alpha(m-\bar m)}  = \delta_{m,\bar m}~.
\ee
We then use the variable $\bar m$ for the right-movers and rewrite the divergent part, by expanding the exponents and collecting terms in the powers of the nome $q$: 
\begin{multline}
Z^{div}_{BTZ}(r_\pm) = \frac{2r_+}{ |\eta(\tau)|^6}
\int_{-\infty}^{+\infty} ds~
 ~q^{\frac{s^2}{k_b-2}-\frac{k_bw^2}{4}(r_- -i r_+)^2}~\bar q^{\frac{s^2}{k_b-2}-\frac{k_bw^2}{4}(r_- + i r_+)^2}\cr
\sum_{w,m,\bar m}\int_0^1 ~d\alpha ~e^{2\pi im(\alpha-\frac{ i k_b w}{2} r_+r_-) }e^{-2\pi i\bar m(\alpha  +\frac{ i k_b w}{2} r_+r_-) }~ 
\delta(\bar{U}_{m,w}) \delta( U_{m,w})
\end{multline}
We  dualize the $(m, \bar m)$ variables by using the  formula: 
\begin{equation}
  \frac{2\pi}{r_- - i r_+}\sum_m \delta(m-w\tau
    )\, e^{2\pi i \tilde\alpha m}
    = \frac{1}{r_- - i r_+}\sum_n e^{-2\pi i w \tau(n-\tilde\alpha)}~.
\nonumber
\end{equation}
For the case at hand we have 
$\tilde\alpha = \alpha - \frac{i k_b w}{2}r_+r_- $.
We have a similar calculation for the right-movers giving rise to an independent dual variable $\bar{n}$. The variable $\alpha$ remains shared between left- and right-movers. We find:
\begin{multline}
Z^{div}_{BTZ} = \frac{ r_+}{2\pi^2(r_+^2+ r_-^2)} 
\int_{-\infty}^{+\infty}ds~   
\sum_{w,n,\bar{n}} 
q^{\frac{s^2}{k_b-2}- \frac{k_bw^2}{4} (r_- - i r_+)^2} \bar{q}^{\frac{s^2}{k_b-2}- \frac{k_bw^2}{4} (r_- + i r_+)^2}
\\ 
\frac{1}{|\eta(\tau)|^6}
\int_0^1 d \alpha 
e^{ -2 \pi i w \tau ({n-\alpha+i\frac{kw}{2} r_+ r_-})  }
e^{ 2 \pi i w\bar\tau  ({\bar{n}- \alpha-i \frac{kw}{2} r_+ r_-}) }
\, . 
\end{multline}
Once more collecting the powers of the nome $q$, we obtain 
\begin{multline}
Z^{div}_{BTZ} = \frac{ r_+}{2\pi^2 (r_+^2+ r_-^2)} \frac{1}{|\eta(\tau)|^6}
\int_{-\infty}^{+\infty}ds~\int_0^1 d \alpha   
\\ 
\sum_{w,n,\bar{n}} q^{\frac{s^2}{k_b-2}-w(n-\alpha + \frac{i k_b w}{2}r_+r_-)
- \frac{k_bw^2}{4} (r_- - i r_+)^2} \bar{q}^{\frac{s^2}{k_b-2}-w(\bar n-\alpha - \frac{i k_b w}{2}r_+r_-)
- \frac{k_bw^2}{4} (r_- + i r_+)^2}
\, . 
\end{multline}
We recognize the expected shifts in the generalized $J^3$ eigenvalues of the continuum sector characters, respecting the quantization condition of the angular momentum \eqref{AngMomEucl}.  In summary, we have succeeded in identifying a modular invariant pole contribution with an integral over twisted sector continuous characters. In the process, we have glued the  continuous characters consistently with angular momentum quantization. It provides an indication of how one can consistently identify Lorentzian long winding strings that respect angular momentum quantization in the Euclidean black hole background. These would provide a quantum counterpart to the long winding strings solutions based on space-like geodesics, discussed in section \ref{Classical}.

\subsubsection*{Discussion}

Our analysis of the Euclidean partition function and its Lorentzian representation theory content is illuminating but it remains partial. A more careful treatment of the expansion is necessary as well as a more rigorous treatment of all poles. 
The techniques that were used for thermal $AdS_3$ were based on the elliptic spectral flow property of affine $sl(2)$ representations. The hyperbolic nature of the energy and angular momentum generators in the Lorentzian BTZ black hole render a direct extension of these techniques  challenging. 

We found the angular momentum quantization equation (\ref{EuclideanQuantizationOfAngularMomentum}). This quantization condition is more readily interpreted in the Lorentzian where it  corresponds to the quantization of a combination of hyperbolic eigenvalues. In the Euclidean, we can propose that if one defines Lorentzian BTZ amplitudes by analytic continuation from the Euclidean $H_3^+$ model, that the poles that capture the physics of the discrete modes reside at the complex value of the parameter $j$ given by the Euclidean quantization rule (\ref{EuclideanQuantizationOfAngularMomentum}).

In our interpretation of the partition function, we tend towards an interpretation that matches the Lorentzian. In anti-de Sitter space, as in ordinary quantum field theory, this entails insisting on a spectrum of continuous energy eigenvalues, despite the fact that they are quantized in Euclidean thermal $AdS_3$. In the case of the black hole space-time, we similarly allowed for continuous energy eigenvalues in the Lorentzian black hole interpretation of the Euclidean partition function.

We hope our analysis will be a useful stepping stone to reach a more rigorous treatment of the partition function and its Lorentzian reading.

\section{Conclusions}
\label{Conclusions}
Building on earlier work \cite{Natsuume:1996ij,Hemming:2001we,Troost:2002wk,Hemming:2002kd,Rangamani:2007fz,Berkooz:2007fe,Mertens:2014nca,Mertens:2015ola}, we have further analyzed the dynamics of fundamental strings in three-dimensional anti-de Sitter black holes with Neveu-Schwarz-Neveu-Schwarz flux. We have stressed a physical ambiguity in the definition of the model and have shown how it is resolved in the Lorentzian Wess-Zumino-Witten
model. We have computed explicit expressions for the classical energy of winding strings based on time-like or space-like geodesics. We supplemented the analysis with a direct analysis in terms of the Nambu-Goto plus flux action of a long string probe. Our results for the classical dynamics of twisted strings are more explicit then those hitherto obtained.

We have also analyzed the Euclidean coset partition function and have shown that it should be interpreted directly as an orbifold conformal field theory partition function. While the Euclidean anti-de Sitter partition function contains two  fugacities corresponding to the temperature and the chemical potential for angular momentum, the Euclidean black hole partition function has two parameters that are fixed by the mass and angular momentum of the black hole.  Moreover, we partially identified the quantum spectral content of the partition function in this spirit. While our analysis is preliminary, we could identify the analytically continued discrete part of the Lorentzian spectrum as well as a modular invariant contribution from the continuous representations.  

It is clear that much further work is needed before declaring that fundamental string perturbation theory is well-defined and understood on the three-dimensional black hole backgrounds under consideration. Concrete, important and  intriguing questions remain largely open. For example, the hyperbolic orbifold may induce divergences in perturbation theory, or may be well-defined through a specific analytic continuation procedure starting from the Euclidean coset model. A more careful decomposition of the partition function may teach us valuable lessons about the perturbative spectrum. The definition of the background in terms of microscopic objects in string theory may inform us how to treat the proposed cut-off surface at $r=0$. A long list of open and  solvable questions invites further work.

\section*{Acknowledgements}

We would like to thank Nemani V. Suryanarayana for helpful discussions and our colleagues in general for creating a stimulating research environment.

\appendix

\section{The Euclidean Black Hole Background}
\label{Geometries}
In this section, we briefly review the Euclidean BTZ black hole geometries in various coordinate systems. The review serves as a  reference for statements made in the bulk of the paper. First of all, we recall that the
 Euclidean  black hole, like the thermal AdS geometry, has the topology of a solid torus \cite{Carlip:1994gc,Maldacena:1998bw}. These geometries have a constant negative curvature metric and are locally Euclidean AdS.
There is a choice of which cycle of the boundary torus is filled in and 
there are a number of inequivalent choices. The Euclidean AdS and the black hole choices are related by an S-duality transformation on the boundary torus \cite{Carlip:1994gc,Maldacena:1998bw}.

 \subsection{The Metric on the Euclidean Black Hole}
\label{BTZmetriccoordinates}
More concretely, starting from the Lorentzian black hole metric   (\ref{BTZMetric}), we analytically continue the inner horizon $r_-$ and time coordinate  $t$ as $r_- \rightarrow i r_-$ and $t=-it_E$ to obtain the  Euclidean black hole metric:
\be
ds^2 = \frac{r^2}{(r^2-r_+^2)(r^2+r_-^2)}dr^2 + \frac{(r^2-r_+^2)(r^2+r_-^2)}{r^2}dt_E^2+r^2(d\phi - \frac{r_+r_-}{r^2} dt_E)^2~.
\ee 
We demand regularity of the Euclidean black hole metric at the outer horizon where the Euclidean time circle pinches off. Along with the usual periodicity condition on the $\phi$ coordinate we therefore have the identifications:
\begin{align}
\label{identifone}
(t_E, \phi) &\sim (t_E +  \frac{2\pi r_+}{r_+^2+ r_-^2}~, \phi +\frac{ 2\pi r_-}{r_+^2 + r_-^2})\, ,  \qquad \text{and} \qquad
(t_E, \phi) \sim (t_E, \phi + 2\pi)~.
\end{align}
In the main text we used Poincar\'e coordinates to describe the Euclidean black hole, in terms of which the solutions to the classical equations of motion were easily written down. These are related to the $(r, t_E, \phi)$ coordinates by the change of variables:
 \begin{align}
 \label{Poincaretortphi}
 \gamma &=\sqrt{\frac{r^2-r_+^2}{r^2+r_-^2}}e^{(r_- - i r_+)(i\phi - t_E)}~,\quad \bar\gamma =\sqrt{\frac{r^2-r_+^2}{r^2+r_-^2}}e^{(r_- + i r_+)(-i\phi - t_E)}\cr
 \xi &= -\frac12\log\left( \frac{r_+^2+r_-^2}{r^2+r_-^2}\right) - r_+ \phi + r_- t_E \, .
 \end{align}
 The metric in Poincar\'e coordinates takes the form: 
\be
ds^2 = d\xi^2 + e^{2\xi}\, d\gamma\, d\bar\gamma~.
\ee
The coordinate identifications  \eqref{identifone} become:
\begin{align} 
(\xi, \gamma, \bar \gamma) \sim (\xi, e^{2\pi \ii}\gamma, e^{-2\pi i }\bar\gamma)~, \quad \text{and} \quad (\xi, \gamma, \bar \gamma) \sim (\xi -2\pi r_+, e^{2\pi \ii (r_- - i r_+)}\gamma, e^{-2\pi \ii (r_- + i r_+)}\bar \gamma) ~.
\nonumber
\end{align}
One of the identifications is trivialized in these coordinates. 
 
\subsection{The Determination of the Partition Function} \label{DeterminationPartitionFunction}
 
In the evaluation of the one loop amplitude for string theory on the black hole background, it turns out to be useful to work in a different set of coordinates that allows one to render the path integration free \cite{Gawedzki:1990ji}. These new coordinates are obtained from the $(\xi, \gamma, \bar\gamma)$ Poincar\'e coordinates by a  rescaling:
\be
v = e^{\xi }\, \gamma~,\quad \bar v = e^{\xi}\, \bar\gamma~,\quad \varphi = -\xi~. 
\ee
 By composing the maps one finds a direct relation between these and the $(r,t_E,\phi)$  coordinates:
\begin{align}
v 
&=\sqrt{\frac{r^2-r_+^2}{r_+^2+r_-^2}}~ e^{i r_+ t_E+i r_- \phi}~,\qquad \bar v 
=\sqrt{\frac{r^2-r_+^2}{r_+^2+r_-^2}}~ e^{-i r_+ t_E -i r_- \phi}~
\nonumber \\
\varphi 
&= \frac12\log\left( \frac{r_+^2+r_-^2}{r^2+r_-^2}\right) + r_+ \phi - r_- t_E \, .
\label{vvbartorphi}
 \end{align}
%
In terms of these  coordinates the metric takes the form
 \begin{align}
 ds^2 
 = d\varphi^2 + (dv + vd\varphi)(d\bar v + \bar v d\varphi)~,
 \end{align}
 and the identifications \eqref{identifone}  become: 
\begin{align}
(\varphi, v, \bar v) &\sim (\varphi, e^{2\pi i} v, e^{-2\pi i} \bar v)~ \, \qquad \text{and} \qquad 
(\varphi, v,  \bar v) \sim (\varphi + 2\pi r_+, e^{2\pi i r_-} v, e^{-2\pi i r_-} \bar v)~,
\label{identiftwo}
\end{align}
The identification due to regularity at the outer horizon becomes trivial in these coordinates. At the same time, the periodicity of the original angular coordinate in the black hole turns into a non-trivial identification.
The vector field that generates the translation in the identified direction can be expressed as:
 \be
 \frac{\pa}{\pa\phi} = r_+ \frac{\pa}{\pa\varphi} + i\, r_-\left(v\frac{\pa }{\pa v} - \bar v\frac{\pa}{\pa\bar v} \right)~.
 \ee
Rewriting the BTZ black hole metric in terms of the $(\varphi, v, \bar v)$ coordinates makes it possible to fluently write down the Euclidean partition function. The metric has the same form as in thermal $AdS_3$. The  identifications of the coordinates are also as in the Euclidean thermal $AdS_3$ space, provided one makes an  analogy between  the parameters of the thermal $AdS_3$ space and the background mass and angular momentum of the black hole:
\begin{align}
\label{adsBHsub}
   \beta  \longleftrightarrow  2 \pi r_+ ~,\qquad
 \beta \mu  \longleftrightarrow  2 \pi r_- \, .
\end{align}
Given the calculation of \cite{ Gawedzki:1990ji, Maldacena:2000kv} for the partition function of thermal $AdS_3$, the partition function of the conformal field theory on the Euclidean black hole background is then  obtained from that of thermal $AdS_3$ by using the substitutions  \eqref{adsBHsub}. One finds the result
\begin{equation}
Z_{BTZ} = \frac{r_+ \sqrt{k_b-2}}{ \sqrt{\tau_2}}
\sum_{w,m} \frac{e^{-\pi \frac{k_b}{ \tau_2} r_+^2 |m-w \tau|^2 + \frac{2 \pi}{\tau_2} Im (\bar{U}_{m,w})^2}}{|\theta_1(\bar{U}_{m,w},\tau)|^2} \, ,  \label{BTZPF}
\end{equation}
where the holonomy $\bar{U}$ is given by
\begin{equation}
\bar{U}_{m,w} = 
(r_- - ir_+) (m-w\tau) \,. 
\end{equation}
The spectral content of the partition function in the Euclidean and its analytic continuation to the Lorentzian is analyzed in the bulk of the paper.

\section{The Energy of a Twisted Short  String}
\label{generalprobe}

In this section we show that the energy of a twisted short string in the background of a rotating black hole computed  in conformal gauge matches the energy of a string calculated using the static gauge in a Nambu-Goto action. We first of all write down the  Nambu-Goto plus flux  action in static gauge:
\be
t= \tau~, \quad \phi = w \sigma~.
\ee
The action for the twisted or winding string then takes the form 
\be
S = k \int dt d\phi \left( w r^2 - 
\sqrt{
w^2(r^2 - r_+^2)(r^2 - r_-^2) +r'^2 
-\frac{(wr^2 \dot r^2+r_+r_-r')^2}{(r^2 - r_+^2)(r^2 - r_-^2)}
}
\, \right)~.
\ee
To compute the energy of the string, we work near the turning point, at which $\dot r=0$, we have $r'=0$ and $r= r_{\text {max}}$, the maximum radial extent of the string. Then, the energy of the string is entirely given by the potential energy:
\begin{align}
E_{\text {string}} &= k w\left( \sqrt{(r_{\text {max}}^2 - r_+^2)(r_{\text {max}}^2 - r_-^2) } - r_{\text {max}}^2\right)~.
\label{probestringenergyexp}
\end{align}
We now compare this with the energy calculated from the orbifold Wess-Zumino-Witten model \eqref{Estringoriginal},
\be
E_{\text {string}} = \frac{h+\bar h}{w} - \frac{k}{2w}m^2 - \frac{kw}{2}(r_+^2 + r_-^2)~,
\ee
where the on-shell conditions fix $m$ and the particle angular momentum $L$ to be (see equation \eqref{msoln})
\be
m = -\widetilde{E} w + \sqrt{ \frac{2h+2\bar h}{k} +w^2(\widetilde{E}^2 + r_+^2 + r_-^2)}~,\quad L = \frac{\bar h- h}{kw} + r_+r_-w~.
\ee
Recall here that $\widetilde{E} = \frac{E}{m}$, the rescaled energy of the particle. 
To compare with the Nambu-Goto calculation which did not assume any internal directions in the target space, we set $h=\bar h=0$, which leads to
\be
\label{msimplersoln}
m=w\left(-\widetilde{E} +\sqrt{\widetilde{E}^2 + r_+^2 + r_-^2}\right)~, \quad L = r_+r_-w~.
\ee
Substituting these equalities into the expression for the energy we obtain
\be
\label{Estringprobesub}
E_{\text {string}} = -kw (r_+^2 + r_-^2 + \widetilde{E}^2)+ kw\widetilde{E} \sqrt{(r_+^2 + r_-^2 + \widetilde{E}^2)}~. 
\ee
To compare the two expressions for the energy of the string, one needs to relate the dimensionless energy $\widetilde{E}$ with the maximal radial extent $r_{\text {max}}$ of the geodesic. The relation can be read off from the geodesic to be 
\be
r_{\text {max}}^2 = \frac{\alpha+\sqrt{\alpha^2 + 4m^2\beta}}{2m^2} ~,
\ee
with $\alpha$ and $\beta$ given in equation \eqref{alphabetadefn}. In those expressions we set $E=m \widetilde{E}$ and $L=r_+r_-w$, substitute for the particle mass $m$ using equation \eqref{msimplersoln} and find that $\beta=0$. Therefore, we have 
\begin{align}
r_{\text {max}}^2 
& =\frac{\alpha}{m^2} =\widetilde{E}^2+ r_+^2+r_-^2  - \frac{r_+^2r_-^2}{(r_+^2 + r_-^2)^2}\left(\widetilde{E} +\sqrt{\widetilde{E}^2 + r_+^2 + r_-^2}\right)^2. 
\label{rmaxExpression}
\end{align}
Inverting the expression, we obtain 
\be
\widetilde{E} = \frac{1}{r_+^2-r_-^2}\left(r_+^2\sqrt{r_{\text {max}}^2-r_+^2} + r_-^2\sqrt{r_{\text {max}}^2-r_-^2} \right)  .
\label{Etildermax}
\ee
Using the relations (\ref{rmaxExpression}) and (\ref{Etildermax})
between the maximal radius $r_{\text {max}}$ and the rescaled particle energy $\widetilde{E}$, one can prove that the energies \eqref{probestringenergyexp} and \eqref{Estringprobesub} obtained from the Nambu-Goto and the conformal gauge approach agree.

\bibliographystyle{JHEP}

\end{document}